\documentclass{article}
\usepackage{authblk}
\usepackage{booktabs}
\usepackage[inline]{enumitem}
\usepackage[a4paper, margin=1in]{geometry}
\usepackage[utf8]{inputenc}
\usepackage[T1]{fontenc}
\usepackage[numbers]{natbib}
\usepackage{pgfplots}
\usepackage[a4paper, margin=1in]{geometry}
\usepackage[hyphens]{url}
\usepackage[english]{babel}
\usepackage{caption}
\usepackage{pgf-pie}
\usepackage{tikz}
\usepackage{pgfplots, pgfplotstable}

\usetikzlibrary{arrows, positioning, shadows}
\definecolor{myred}{HTML}{ef476f}
\definecolor{myblue}{HTML}{118ab2}
\definecolor{myyellow}{HTML}{ffd166}
\definecolor{mygreen}{HTML}{06d6a0}
\definecolor{mydarkblue}{HTML}{073b4c}
\usepackage{amsmath}
\usepackage{graphicx}
\usepackage{todonotes}
\usepackage{caption}
\usepackage{subcaption}

\title{What's inside a node? Malicious IPFS nodes under the magnifying glass}

\providecommand{\keywords}[1]{\textbf{\textit{Index terms---}} #1}
\begin{document}
\author[1]{Christos Karapapas}
\author[1]{George C. Polyzos}

\author[2,3]{Constantinos Patsakis}
\affil[1]{Athens University of Economics and Business, Greece}
\affil[2]{Department of Informatics, University of Piraeus, 80 Karaoli \& Dimitriou str., 18534 Piraeus, Greece}

\affil[3]{Information Management Systems Institute of Athena Research Centre, Greece}

\date{}
\maketitle

\begin{abstract}
    
   InterPlanetary File System~(IPFS) is one of the most promising decentralized off-chain storage mechanisms, particularly relevant for blockchains, aiming to store the content forever, thus it is crucial to understand its composition, deduce actor intent and investigate its operation and impact. Beyond the network functionality that IPFS offers, assessing the quality of nodes, i.e. analysing and categorising node software and data, is essential to mitigate possible risks and exploitation of IPFS. To this end, in this work we took three daily snapshots of IPFS nodes within a month and analysed each node (by IP address) individually, using threat intelligence feeds. The above enabled us to quantify the number of potentially malicious and/or abused nodes. The outcomes lead us to consider using a filter to isolate malicious nodes from the network, an approach we implemented as a prototype and used for assessment of effectiveness.

\end{abstract}
\keywords{InterPlanetary File System, Web3 Security}

\section{Introduction}
Web 1.0 is known as the \textit{read-only} web. For many years the Word Wide Web had an informative and educative role presented through static content. Few people generated content that was read by many people. 
From the constant interaction as well as the expanding familiarity that users had with the web, the number of users who also wanted to create content grew. Thus, the need for a more participative web arose, giving birth to Web 2.0. The latter, despite its shortcomings, is massively adopted. Single points of failure due to security incidents and control from a single organisation along with privacy violations for, e.g.  marketing purposes, by centralised data storage facilities have been two of the most thorny issues in Web 2.0 for years.

Lately, there has been a lot of discussion around Web3. One of the pillars of Web3 is the decentralisation of the web to allow users to regain control over their data 
and selectively share and monetise the information they create.
An integral part of attaining these goals are distributed ledger and blockchain technologies, and token-based economics \cite{cook2020spatial,murray2022promise}. Web3 promises to offer decentralised services, meeting the needs of the Internet of Things~(IoT) era and introducing a financial aspect of the web-user relationship through cryptocurrencies.

Web3 is considered to consist of different stacks, and these, in turn, of different protocols that cooperate with each other in order to provide services to the user. Some of them are data storage, domain name resolution, decentralised identities or, at a higher level, social media, gaming and marketplaces. All these different protocols, as well as the bridges between them, are still in their making; thus, their shortcomings may be exploited by attackers or utilised  
for malicious purposes. Indeed, 
ransomware and dark web marketplaces use cryptocurrencies to make siphon their payments, while blockchains and IPFS are used for the coordination of malware as C2 servers or to store malicious payloads.  

In this work, we focus on distributed data storage and, more specifically, the InterPlanetary File System~(IPFS), a cornerstone of the decentralised storage component of Web3. IPFS claims to have $2$ million unique weekly users~\footnote{\url{https://decrypt.co/resources/how-to-use-ipfs-the-backbone-of-web3}}, and it has certainly caught the eye of the scientific community, as reflected by a total of more than 1160 papers found on Scopus 
with the search term ``IPFS'' in title and/or abstract. As IPFS is a collection of sub-protocols, it can be exploited by malicious users in a variety of ways. Immutability and decentralisation create a very dangerous mix that can be abused in various ways \cite{casino2019immutability}. Karapapas et al. \cite{raas} illustrated how cybercriminals could exploit IPFS to set up an anonymous malware C2 facility alongside smart contracts. Patsakis et al. \cite{playground} showed that it could be abused to provide a robust malware C2 server infrastructure. Moreover, it is known to have been utilised by the Storm botnet\cite{ipstorm} while new evidence has come to light linking IPFS to phishing~\footnote{\url{https://www.trustwave.com/en-us/resources/blogs/spiderlabs-blog/ipfs-the-new-hotbed-of-phishing}}.

To this end, we aim to unravel the structural elements of the IPFS network, and the nodes, focusing on suspicious activity. Initially, we crawl the IPFS network to enumerate it and make the first contact with the nodes. Following that, we collect intelligence from different sources regarding the aforementioned nodes. Moreover, we collect the exchanged data by nodes and analyse them to have a deeper understanding of the consistency of the network. Finally, we try to determine the extent of possible abuse of IPFS for copyright infringement.

The remainder of this paper is structured as follows: In Section 2, we present the required background information and overview of technologies. In Section 3, we present related research regarding IPFS monitoring. Section 4 focuses on nodes, presenting our data collection methodology and our findings regarding the nodes that constitute the IPFS network. 
Then, Section 5 goes a step higher in terms of abstraction, focusing on the content stored in IPFS. In Section 6, we discuss possible countermeasures to isolate malicious nodes. Finally, in Section 7, we summarise our findings and contributions, discussing possible future research directions.

\section{Background}
\subsection{IPFS}
InterPlanetary File System~(IPFS) \cite{benet2014ipfs} is a peer-to-peer file-sharing system, consisting of many novel technologies, aiming to achieve decentralised data storage and low latency file distribution. Some of its main goals are to foster censorship circumvention and to avoid a single point of failure. E.g., in 2017 it was utilised to disseminate and store data regarding the Catalan independence referendum\footnote{\url{https://edri.org/our-work/no-justification-for-internet-censorship-during-catalan-referendum/}} when the Spanish government attempted to censor it. 

Contrary to traditional file systems, in IPFS files are addressed by their content and each one is assigned a unique content ID~(CID). One of the main IPFS components is \texttt{libp2p}\cite{protocol2021libp2p}, an umbrella term for many underlying network protocols. IPFS uses Distributed Hash Table~(DHT), a highly scalable coordinator of data lookup
among the different nodes. libp2p provides IPFS with the \texttt{KAD-DHT}, a Kademlia \cite{maymounkov2002kademlia} variant. The latter is responsible for storing three types of mappings:
\begin{enumerate*}
    \item Provider Records, i.e., what content is hosted by whom, 
    \item Peer Records, i.e., who (PeerID) has what address and finally, 
    \item InterPlanetary Name System~(IPNS) records, i.e., static names pointing to varying data. 
\end{enumerate*}
Another noteworthy component is BitSwap, which is a data-exchanging protocol based on \texttt{want-have content} and \texttt{have content} messages\cite{bitswap}. Moreover, Merkle DAG, a combination of Merkle Tree and Directed Acyclic Graph~(DAG), is used to certify that the data exchanged are unique and IPFS does not store any duplicates. Finally, users can have access to files stored on IPFS, through HTTPs, by visiting public gateways. Public gateways have been provided not only by Protocol Labs, which is the main developer of IPFS but also by various companies embracing Web3, like Cloudflare, Pinata, etc~(\url{https://ipfs.github.io/public-gateway-checker/}). As of July $2022$, i.e., \texttt{v0.14}, the implementation of IPFS is known as \texttt{Kubo}~\footnote{\url{https://github.com/ipfs/kubo/blob/master/docs/changelogs/v0.14.md}}.

\section{Related Work}
P2P networks have been of interest to the scientific community for many years, and while their popularity fluctuates, they have never been outdone. In recent years, the advent of cryptocurrencies and blockchain technology has brought them back into the limelight. Thus, while P2P node profiling has been extensively studied in the past, research in the context of Web3 is minimal. Web3 is in a very early phase and its decentralised components are still under heavy development. Hence, the current research regarding its nodes is still in its infancy.
Henningsen et al. in~\cite{crawl} make one of the first attempts to explore the IPFS network. Adopting a hybrid design, passively and actively, they aim to enumerate the IPFS network and profile its nodes. The authors note that the overlay network outperforms the overlay induced by buckets. Furthermore, they observe that an overwhelming percentage of nodes, i.e. 94\%, did not react to the authors' attempt to connect to them. The reason this happens is twofold. The first is because many nodes are behind NAT and thus advertise their local IP address. The second is that a large portion of users uses IPFS in an opportunistic way, therefore their footprint remains in buckets for longer than they remain online and connected.

Recently,  researchers discovered a botnet hiding in the IPFS ecosystem \cite{ipstorm}. The latter, named InterPlanetary Strom~(IPStorm) and estimated size of 9000 devices, utilises IPFS at multiple levels. Initially, the researchers found that it uses the libp2p DHT to discover nodes. Bots identify each other with the attribute \texttt{Agent Version}: ``storm''. In addition, the botnet utilises the Pub/Sub protocol as a communication channel over specific topics. Finally, the botnet uses IPFS to share files so that it can be updated to a newer version.

Trautwein et al. \cite{trautwein2022design} further to providing a basic guide of IPFS' design, they collected data from three different sources to shed light on various metrics related to IPFS performance. Initially, they crawled the IPFS network to gather information about peers. Among the conclusions drawn is that IPFS nodes are geographically distributed in 152 countries, yet more than 50\% are located in just two countries, US and China. Furthermore, more than 50\% of the IPs are covered by five automated systems, yet only 2.3\% of the nodes are in some cloud infrastructure. The last insight extracted from this dataset is that the IPFS network suffers from high rates of churn, with 87.6\% of peers having an uptime of less than 8 hours. Finally, the authors wanted to study the time performance in downloading data. To this end, they experimented with different AWS regions and recorded the download duration from the data they produce each time. In 50\% of the cases, the download took less than 3s, and in 90\% of the cases, less than 4.5s.

\section{Profiling IPFS nodes}
\subsection{Data collection methodology}
To enumerate the IPFS network we used the IPFS Crawler~\cite{crawl}.
The IPFS crawler is a tool written in Go and is based on libp2p (v0.11.0).
Acting as a Kademlia node the crawler uses precomputed keys to extract all the entries from most buckets for every node it encounters.
In essence, it invokes \textsc{FindNode} actions repeatedly using the appropriate precomputed keys.
Finally, the crawler produces two files: (i)~a JSON file storing the tuple \texttt{$<$PeerID, multiaddress, agent, reachability$>$} for every distinct node met, and (ii)~a CSV file containing all the pairs of connected nodes.

We conducted a series of consecutive crawls. Initially, the crawls were performed iteratively, every ten days 
during the period from March to April 2022. Each crawl series spanned over a day (24h) totalling about 360 crawls in a row per day.  From the data in the JSON file, for each PeerID we extracted the IP addresses. Each IPFS node maintains an address book retaining information for the nodes it encounters. If any of the encountered nodes advertises a new address, then it is appended in the address book for reachability purposes. As a result, a single PeerID may correspond to more than one IP address. We studied each different address considering it as a unique node. Moreover, nodes behind a firewall or NAT use \texttt{p2p-circuit}, a libp2p relay transport protocol, to avoid connectivity barriers. In essence, these nodes advertise addresses through relay nodes. As a consequence, they do not reveal their real IP address but the IP address of the relay. The aforementioned peers as well as those which advertise only local IP addresses are excluded from our analysis. Clearly, the absence of such IP addresses prevents us from studying or fingerprinting the corresponding hosts.

\subsection{Node Profiling}
In this section, we present general information regarding the IPFS network and its nodes. We should mention that in the following findings, every different IP address is considered a different node. Although, we found that unique Peer IDs advertise multiple IP addresses since our study focuses on the ``fabric'' of the IPFS network. Thus, we want to enumerate and analyse every different IP address.




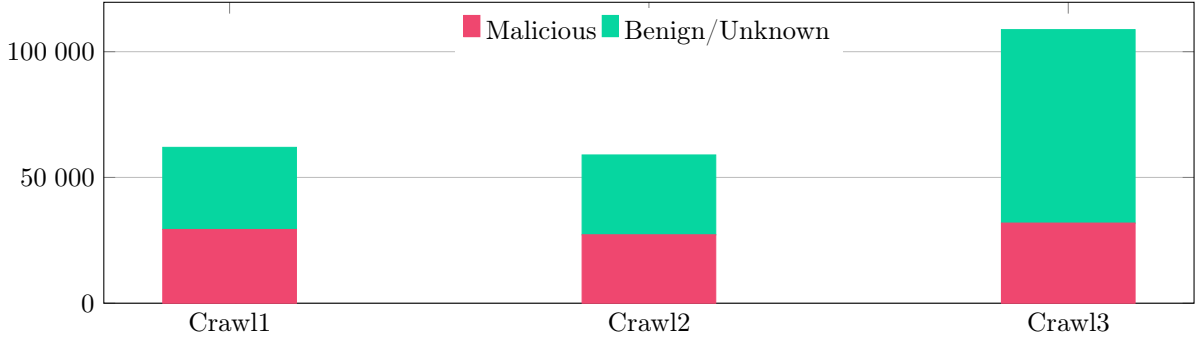
\begin{figure}[th!]
\centering
    \begin{tikzpicture}
    \pgfplotsset{width=\textwidth,compat=1.8}
    \pgfplotsset{every axis/.append style={
        scaled y ticks = false, 
        scaled x ticks = false, 
        y tick label style={/pgf/number format/.cd, fixed, fixed zerofill,
                            int detect,1000 sep={\;},precision=3},
        x tick label style={/pgf/number format/.cd, fixed, fixed zerofill,
                            int detect, 1000 sep={},precision=3}
    }
}
      \begin{axis}[ymajorgrids = true,
        ybar stacked, ymin=0,  
        bar width=50pt,
        enlarge x limits=0.15,
        legend columns=-1,
        legend style={draw=none,at={(0.5,0.98)},anchor=north},
        symbolic x coords={Crawl1,Crawl2,Crawl3,Day1,Day2,Day3,Day4,Day5,Day6,Day7},
        xtick=data,
        nodes near coords align={anchor=north},
        every node near coord/.style={},height=.35\textwidth,
      ]
      \addplot [myred,fill=myred] coordinates {
    ({Crawl1},29313)
    ({Crawl2},27201)
    ({Crawl3},31855)
    };
      \addplot [mygreen,fill=mygreen] coordinates {
    ({Crawl1},32647)
    ({Crawl2},31739)
    ({Crawl3},76901)
    };
    \draw [dashed] (1.5,0) -- (1.5,1);
      \legend{Malicious,Benign/Unknown}
      \end{axis}
  \end{tikzpicture}
  \caption{Malicious nodes per crawl.}
  \label{fig:malicious_nodes}
\end{figure}
\begin{figure}[thb]
    \centering
    \begin{subfigure}[b]{0.32\textwidth}\centering
        \resizebox{\textwidth}{!}{
        \begin{tikzpicture}
          \tikzset{venn circle/.style={draw,circle,minimum width=6cm,fill=#1,opacity=0.50,draw=none}}
          \node [label] at (0,-3.5) {\Large{\textbf{Crawl 1}}};
          \node [venn circle = myred] (A) at (0,0) {\Large{\textbf{61960}}};
                    \node[label] at (2,7) {\Large{\textbf{Crawl 2}}};
          \node [venn circle = myblue] (B) at (60:4cm) {\Large{\textbf{58940}}};
          \node [venn circle = myyellow] (C) at (0:4cm) {\Large{\textbf{108756}}};
          \node [label] at (4,-3.5) {\Large{\textbf{Crawl 3}}};
          \node [left] at (barycentric cs:A=1/3,B=1/2 ) {\Large{\textbf{22347}}}; 
          \node[below] at (barycentric cs:A=1/2,C=1/2 ) {\Large{\textbf{20088}}};   
          \node[right] at (barycentric cs:B=1/2,C=1/3 ) {\Large{\textbf{23130}}};   
          \node[below] at (barycentric cs:A=1/3,B=1/3,C=1/3 ){\Large{\textbf{16783}}};
        \end{tikzpicture}
}
        \caption{\centering{Count of IP addresses per crawl.}}
        \label{fig:venn_intersection}
    \end{subfigure}
    \begin{subfigure}[b]{0.32\textwidth}\centering
         \resizebox{\textwidth}{!}{
        \begin{tikzpicture}
          \tikzset{venn circle/.style={draw,circle,minimum width=6cm,fill=#1,opacity=0.5,draw=none}}
          \node [venn circle = myred] (A) at (0,0) {\Large{\textbf{16932}}};
                    \node[label] at (0,-3.5) {\Large{\textbf{Crawl 1}}};

          \node [venn circle = myblue] (B) at (60:4cm) {\Large{\textbf{13415}}};
          \node[label] at (2,7) {\Large{\textbf{Crawl 2}}};
          \node [venn circle = myyellow] (C) at (0:4cm) {\Large{\textbf{15843}}};
                    \node[label] at (4,-3.5) {\Large{\textbf{Crawl 3}}};

          \node[left] at (barycentric cs:A=1/3,B=1/2 ) {\Large{\textbf{3512}}}; 
          \node[below] at (barycentric cs:A=1/2,C=1/2 ) {\Large{\textbf{2290}}};   
          \node[right] at (barycentric cs:B=1/2,C=1/3 ) {\Large{\textbf{3636}}};   
          \node[below] at (barycentric cs:A=1/3,B=1/3,C=1/3 ){\Large{\textbf{5127}}};
        \end{tikzpicture}
}
    \caption{\centering{Count of the malicious IP addresses per crawl.}}
    \label{venn_intersection_mal}
    \end{subfigure}
    \begin{subfigure}[b]{0.32\textwidth}\centering
         \resizebox{\textwidth}{!}{
      \begin{tikzpicture}
\pie[text=pin,color={myblue, myred, mygreen,myyellow,white}]{69/Ubuntu, 19/Windows, 5.5/Asus-WRT, 4.5/Debian, 2/DSM }

\end{tikzpicture}
}
\caption{\centering{Top five OS used by IPFS nodes.}}
    \label{fig:os_stats}
\end{subfigure}
    
    \caption{Crawl statistics.}
\end{figure}
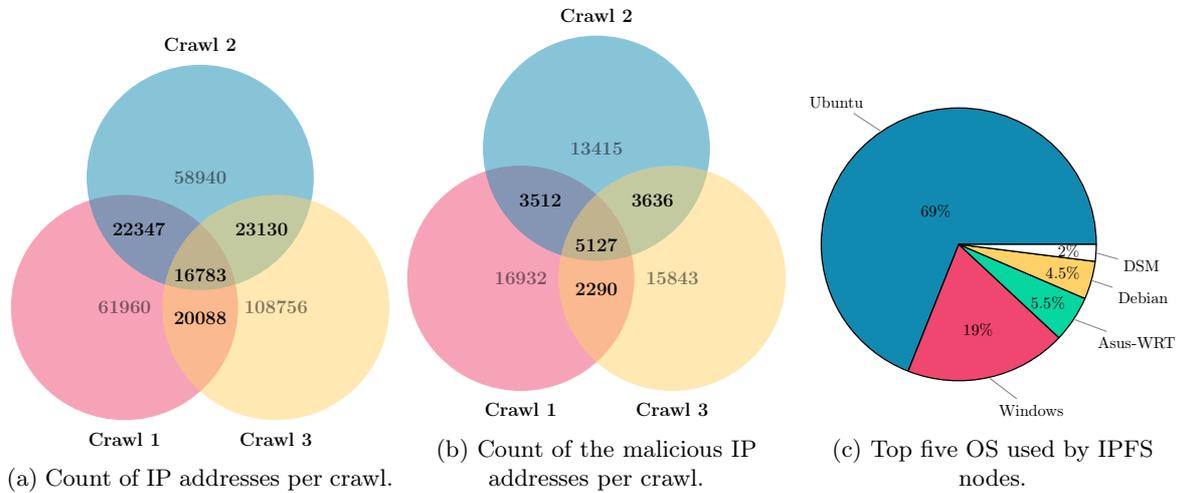

Figure \ref{fig:malicious_nodes} illustrates the nodes per crawl and the count of malicious nodes for which we collected intelligence.
In Figure~\ref{fig:venn_intersection}, the exact results of IP addresses per crawl can be found. Moreover, from the same figure, we can observe that $16783$ were found online in all three crawls. We can assume that the aforementioned nodes were found online at least once a day in the span of the whole month. Given the periodic changes of IPs, we can assume that most of these IPs belong to some infrastructure that has been devoted to constantly working with IPFS.

\begin{figure}[th]
\centering
\begin{tikzpicture}  
\begin{axis}  
[  
    ybar,  scaled y ticks = false, 
        scaled x ticks = false, ymajorgrids = true,
    width=\textwidth, ylabel=\#IPs,
    height=.35\textwidth,    
    ymin=0,
    bar width=5pt,
    xticklabels={{go-ipfs/0.8.0/*.*},{go-ipfs/0.9.0/*.*},{go-ipfs/0.7.0/},ioi,storm,{hydra-booster/0.7.4},{go-ipfs/0.10.0/},{go-ipfs/0.11.0/},{go-ipfs/0.12.0/}
    },
    xtick=data,  
    label style={font=\tiny},
    xticklabel style={xshift=-15pt, rotate=45,font=\small},
    legend style={draw=none,at={(0.25,0.9)},anchor=north},legend columns=-1,
    after end axis/.code={ 
            \draw [ultra thick, white, decoration={snake, amplitude=1pt}, decorate] (rel axis cs:0,1.05) -- (rel axis cs:1,1.05);
        },
    ]  
\addplot [myred,fill=myred] coordinates {(0,11780+1007+480) (1,3641+480) (2,3172) (3,2543) (4,2289) (5,926) (6,1334) (7, 660) (8, 301)
};  
\addplot [myyellow,fill=myyellow] coordinates {(0,6045+1047+446) (1,3250+446) (2,2742) (3,557) (4,1186) (5,845) (6,1145) (7, 514) (8, 270) 
};
\addplot [mygreen,fill=mygreen] coordinates {(0,6469+1007+447) (1,3103+447) (2,2804) (3,821) (4,1253) (5,36040) (6,1156) (7, 503) (8, 295) 
}; 
  \legend{Crawl1,Crawl2,Crawl3}
\end{axis}  
\end{tikzpicture}
\caption{The ten most commonly used agent versions in each crawl. The \texttt{*.*} denotes varying subversions combined.}
\label{fig:user_agents}
\end{figure}
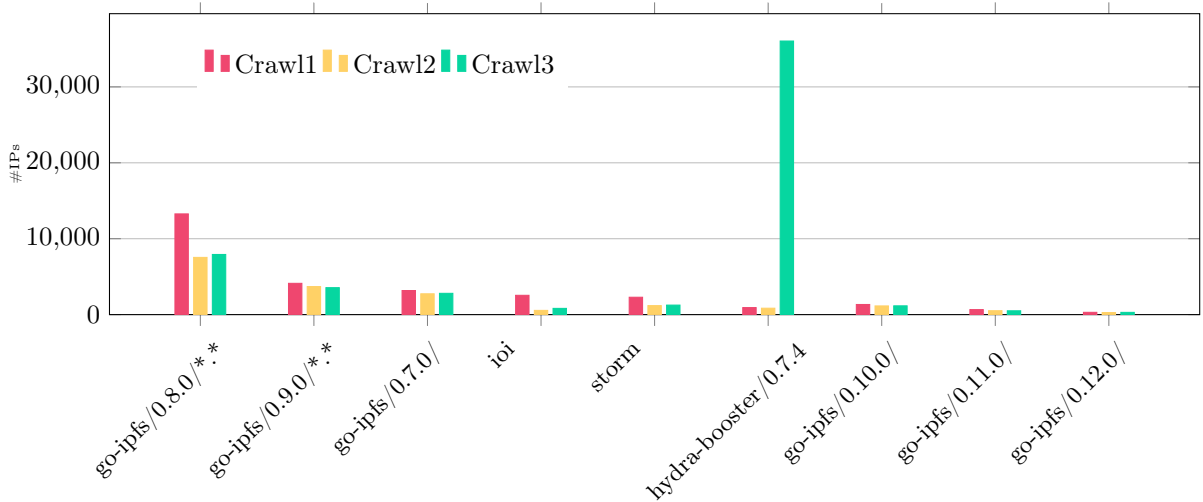

A node's agent version can be an indication of malicious activity. Nodes' agent version is public and advertised, thus, it can act as an identifier for malicious nodes to discover and track each other. The latter is a technique already implemented by ``storm'' agents. Figure \ref{fig:user_agents} illustrates the ten most used agent versions we found in each crawl. We should highlight that the counts depicted correspond to the agents from the nodes we managed to connect to. In each crawl we found 50\%, 61\%, 49\% respectively, unreachable peers, i.e., we found their address stored in the DHT but they were offline. Moreover, IPFS is open-source software; therefore, it is at the user's discretion whether to display the agent version. The latter results are aligned with the ones in \cite{trautwein2022design}. 
In the third crawl we observe that there is an increase in nodes using the agent called \texttt{Hydra Booster}~\footnote{\url{https://github.com/libp2p/hydra-booster/}}. Hydra Booster is a node having many different Peer IDs over a common routing table. It is designed to accelerate IPFS' processes carried out through DHT-like content resolution, routing and discoverability. The existence, as well as the operation of these nodes, brought about an increase in the number of nodes of the third crawl. One of the features of open software, which has been hotly debated lately, is that upgrading to a newer version is at the user's discretion. Observing the crawling results of Figure \ref{fig:user_agents}, one can observe that there are many different software versions running and communicating simultaneously. For example, \texttt{go-ipfs 7.0} was released in July of 2020 while \texttt{go-ipfs 11.0} in August of 2021. Moreover, although the measurements were made in mid-2022, and version \texttt{go-ipfs 12.0} had already been released, we can conclude from the bar charts that the versions which are more widely used are the older ones. In addition, we must mention that agent storm, which has been found in all three crawls with a non-negligible number, is characteristic of the nodes belonging to the IPStorm botnet we have already mentioned.


In what follows, we study the maliciousness of nodes, so we used Virus Total to assess the corresponding IPs. Nevertheless, Virus Total also provides valuable insights regarding the geographic distribution of the various nodes, regardless of whether they are malicious or not. The vast majority of the nodes are located in two countries, namely the United States and China. We notice that our results are aligned  with \cite{trautwein2022design}. 

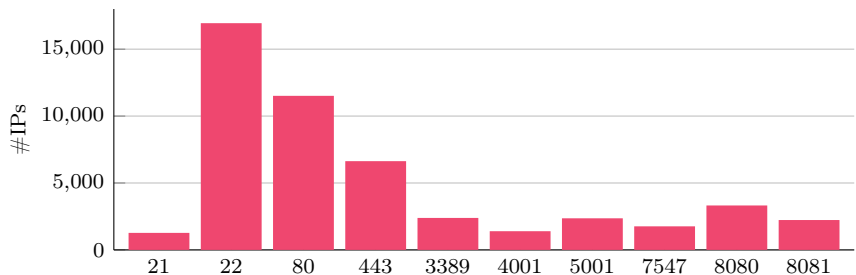
\begin{figure*}[th]
\centering
\begin{tikzpicture}
\begin{axis}[
    ybar stacked,
    xtick=data,
    ymajorgrids,
    axis y line*=none,
    axis x line*=bottom,
    ylabel={\#IPs},
    tick label style={font=\footnotesize},
    label style={font=\footnotesize},
    height=0.3\textwidth,
    bar width=8mm,
    xticklabels={21,22,80,443,3389,4001,5001,7547,8080,8081}, scaled y ticks = false, 
        scaled x ticks = false,  
    ymin=0,
    ymax=18000,
    area legend,
    x=9.5mm,
    enlarge x limits={abs=0.625},
]
\addplot[fill=myred,draw=none] coordinates {(0,1272)(1,16941)(2,11512)(3,6631)(4,2389)(5,1397)(6,2361)(7,1761)(8,3320)(9,2232)};%
\end{axis}
\end{tikzpicture}
\caption{The ten most common ports.}
\label{fig:most_common_ports}
\end{figure*}

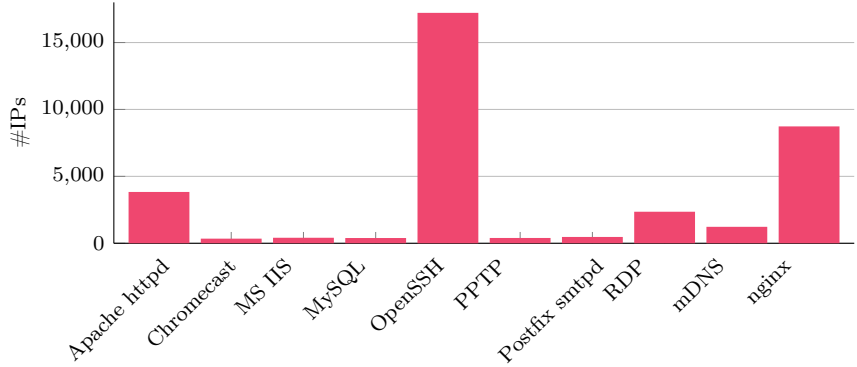
\begin{figure}[h]
\centering
\begin{tikzpicture}
\begin{axis}[
    ybar stacked,
    xtick=data,
    ymajorgrids,scaled y ticks = false, 
    axis y line*=none,
    axis x line*=bottom,
    ylabel={\#IPs},
    tick label style={font=\footnotesize},
    label style={font=\footnotesize},
    height=0.3\textwidth,
    bar width=8mm,
    xticklabels={Apache httpd,Chromecast,MS IIS,MySQL,OpenSSH,PPTP,Postfix smtpd,RDP,mDNS,nginx},  
    ymin=0,
    ymax=18000,
    area legend,
    x=9.5mm,
    enlarge x limits={abs=0.625},
        every node near coord/.append style={font=\tiny},
xticklabel style={xshift=-15pt, rotate=45},
]
\addplot[fill=myred,draw=none] coordinates {(0,3835)(1,346)(2,412)(3,393)(4,17217)(5,397)(6,473)(7,2357)(8,1230)(9,8733)};%
\end{axis}
\end{tikzpicture}
\caption{The ten most commonly used services.}
\label{fig:most_common_services}
\end{figure}

To conduct a more in-depth analysis, we passed the crawling results to intelligence services. Namely, we used Shodan
, a network monitoring tool, to fingerprint each node. Shodan returned intelligence for approximately 40960 unique nodes. Figure~\ref{fig:most_common_ports} illustrates the ten most commonly used ports by the total of nodes we examined. Port 22, the most widely used port by IPs related to IPFS, is typically used for Secure Shell (SSH) connections, which allow users to log in to a host and execute commands remotely. Port 80 is used as the default port for HTTP~(Hypertext Transfer Protocol) traffic, port 8080 is an alternative to port 80 and moreover the default port of the IPFS gateway, and port 443 for HTTPS. Port 3389 is typically used by hosts running Microsoft Remote Desktop Protocol (RDP) to allow remote access to the host's desktop. Finally, port 4001 is used by default for IPFS traffic, but users can also set up a custom port. 
Regarding the operating system running on IPFS nodes, Shodan's results, depicted in Figure~\ref{fig:os_stats}, indicate that the lion's share uses Ubuntu Linux. The next runner-up is Microsoft Windows 10, followed by Debian Linux. The latter is also exhibited by the most used services, Figure \ref{fig:most_common_services}, where most hosts appear to be using SSH as opposed to RDP. Moreover, most of them seem to have a web server (nginx and then Apache).

JARM~\cite{jarm} is an open-source fingerprinting tool that generates a string based on the response of the host to ten TLS packets. JARM is used by the community as a software-wise host clustering tool, therefore it is also eligible to detect malware Command \& Control~(C2). We use JARM strings, extracted from Shodan and Virus Total, to detect any similarities among the different nodes. Finally, we combined them since for the same IP  different services can provide varying information. For 1002 IP addresses, we found information in both services, so we considered both records. The JARMs indicate that there are several clusters of IPs in which servers have the same TLS configuration, which implies that the same entity is behind them. The most common ones are illustrated in Table \ref{tbl:jarm}.

\begin{table}[thb]
\centering
\small
\begin{tabular}{cc}
\toprule
\textbf{JARM} & \textbf{\# IPs}\\
\midrule
\texttt{2ad2ad0002ad2ad00042d42d0000008aec5bb03750a1d7eddfa29fb2d1deea} &2070\\
\texttt{2ad2ad16d2ad2ad22c2ad2ad2ad2adfd9c9d14e4f4f67f94f0359f8b28f532}& 1378\\
\texttt{15d3fd16d29d29d00042d43d000000fe02290512647416dcf0a400ccbc0b6b}&577\\
\texttt{15d3fd16d29d29d00042d43d0000009ec686233a4398bea334ba5e62e34a01} & 562\\
\texttt{15d3fd16d21d21d00042d43d000000fe02290512647416dcf0a400ccbc0b6b} & 489\\
\bottomrule
\end{tabular}
\caption{Most common JARMs.}\label{tbl:jarm}
\end{table}

\subsection{Malicious Activity}
In this section, we investigate the moral character of IPFS nodes, i.e., we examine whether and to what extent there are malicious nodes. To this end, we collect and leverage existing intelligence to create and present their profile. Our goal is to assess the network structure, keeping IPFS users and the related community alert to the existence of malicious activity in the IPFS network. Due to the current IPFS rules, every node maintains several active connections varying from 600 to 900 peers. Thus, we argue that it is very important for each node to know what kind of alignment, i.e. neutral or malicious, the node it interacts with has.

Initially, we leveraged the intelligence provided by two popular services, namely Virus Total (\url{https://virustotal.com/}) and SpamHaus (\url{https://www.spamhaus.org/}), to get a baseline for the reputation and past activity of nodes. SpamHaus uses several methods to find information about an internet resource. It uses sensors in large networks, i.e. a data-sharing community, from which it collects data about network traffic. In addition, SpamHaus deploys honeypots to attract malicious users. Along the same lines as SpamHaus and VT, in addition to monitoring more than 70 anti-malware and IP blocking services, it relies on data generated and shared by an already large community. Both the aforementioned services provide APIs to interact with their knowledge base and generate a JSON formatted output for each request. We combine the extracted output information with the SpamHaus output and we consider malicious those nodes with at least one record in one of the aforementioned services.


Moreover, in Figure \ref{venn_intersection_mal}, we notice that from the $27861$ different IP addresses we encountered during the first crawl, $5126$ of them, $\approx 18\%$ remained online throughout the whole month. The latter indicates that there is a number of nodes that constantly utilise the IPFS network for malicious purposes. Compared to the $16783$ found online in all three crawls, as depicted in Figure ~\ref{fig:venn_intersection}, a significant part of them, i.e., $30.5\%$, are known to be malicious. 
Based on SpamHaus' results, we conclude that the majority of malicious nodes were discovered using the \texttt{DNS Sinkhole} technique. According to this technique, security researchers create, at various levels, a DNS record of a known malicious URL pointing to an address they own, usually a sinkhole server. The gain from applying this technique is twofold: On the one hand, they prevent communication between bot and C2, and on the other hand, researchers can find which computers are infected, i.e. ask to connect to known malicious URLs. 

In Table~\ref{tab:sinkholed_urls},  the five most commonly requested and sinkholed URLs in the number of unique IP addresses are illustrated.
Note that several URLs such as \texttt{differentia.ru}, \texttt{atomictrivia.ru}, \texttt{amnsreiuojy.ru} and \texttt{restlesz.su} are known to be leveraged as C2 by malware. \texttt{disorderstatus.ru} is a relatively newly created domain reported to be mostly used for spamming. To draw deeper conclusions about the URLs, we isolated the \texttt{Top Level Domain}~(TLD) of the different requested URLs. To our surprise, while most requested URLs have a ``.ru'' TLD, this is not reflected among the unique TLDs. On the contrary, we notice that the most commonly encountered is ``.xyz'', a relatively new TLD offering many domains that would traditionally be registered by legitimate users. The fact that they are new and cheap and that traditional domain names are available has led \texttt{xyz} domains to be widely exploited\footnote{\url{https://www.spamhaus.com/resource-center/getting-the-low-down-from-xyz-registry-on-combating-domain-abuse/}\url{https://www.bleepingcomputer.com/news/security/these-are-the-top-level-domains-threat-actors-like-the-most/}}. Given that $11227$ \texttt{xyz} domains are hosted by these addresses makes us conclude that some adversaries use nodes of IPFS for hosting malicious domains in addition to C2 infrastructure.
Tinba a portmanteau of the words Tiny Banker, is a trojan that leverages packet sniffing to determine whether the user visits a bank's webpage. In that case, the trojan tries to steal the keystrokes and sends them to a C2. Nymaim and Ranbyus are well-known trojans, which steal information from the user and consequently send them to a C2. Some of their variants have been found to use domain fluxing to communicate with their orchestrator, and some have been found in DoS attacks. Mirai is used to infect Internet of Things (IoT) devices and turn them into bots that can be used to launch large-scale network attacks. The Mirai botnet was initially discovered in 2016 and was part of various high-profile cyberattacks, including distributed denial-of-service (DDoS) attacks that brought down popular websites and online services. The most frequently displayed campaigns are gathered in Table~\ref{campaigns_count}.

\begin{table}[th]
\centering
\begin{subtable}[h]{0.45\textwidth}\centering
\begin{tabular}{lr}

\toprule
\textbf{URL}      & \textbf{Count} \\ 
\midrule
\texttt{differentia.ru}    & 38681          \\ 
\texttt{disorderstatus.ru} & 15504          \\ 
\texttt{atomictrivia.ru}   & 7049           \\ 
\texttt{amnsreiuojy.ru}    & 5662           \\ 
\texttt{restlesz.su}       & 2180           \\ 
\bottomrule
\end{tabular}
\caption{The five most sinkholed URLs and the number of unique requests.}
\label{tab:sinkholed_urls}
\end{subtable}\quad
\begin{subtable}[h]{0.45\textwidth}\centering
\begin{tabular}{lr}
\toprule
\textbf{Campaigns} & \textbf{Count} \\ \midrule
tinba              & 30019          \\
conflicker         & 22650          \\
nymaim             & 22228          \\
andromeda          & 6403           \\
ranbyus            & 4845           \\
mirai              & 3750           \\
\bottomrule
\end{tabular}
\caption{Malware campaigns with the largest participation from the encountered nodes.}
\label{campaigns_count}
\end{subtable}
\caption{Extroversion of malicious nodes: Which groups do they belong to and what webpages they seek to visit.}
\end{table}
Finally, we studied the JARMs of malicious nodes to better frame our research. As we have already mentioned, we combined knowledge from all intelligence services to produce the results. Notably, among them, we found a cluster of $68$ nodes corresponding to the JARM fingerprint \texttt{15d3fd16d29d29d00042d43d00\\00009ec686233a4398bea334ba5e62e34a01} which is attributed to the notorious \texttt{emotet} botnet.  

As already mentioned, the crawler we used, in addition to information about the nodes encountered, produces an edge list with each pair of connected nodes. Based on this, we constructed a mapping from one PeerID to the several PeerIDs we found connected during the second day. In essence, we built for each peer its buckets expanded to the span of a day. Consequently, we converted the aforementioned mapping to the corresponding IP addresses. This way, we can investigate whether there is a clique between the malicious nodes. The findings indicate that there is no such clique, as the median percentage of malicious nodes in the buckets of a malicious node is 7\%, and the average is 9.5\%. Along the same lines, the median percentage of nodes in the buckets of a benign node is also 7\%, with the average being 9.2\%.

\section{File Investigation}
Despite the processes and functionality IPFS offers through libp2p and its other components, its main purpose is undeniably storage-related. The largest NFT marketplaces use IPFS for the data storage and integrity it provides, while its widespread utilisation has already brought about the need for cooperation with other Web3 layers, such as ENS, which natively offers names corresponding to CIDs. No wonder the increasing popularity has also caught the eye of cyber criminals. A recent research~\footnote{\url{https://blog.talosintelligence.com/ipfs-abuse/}} highlights that the volume of malware samples hosted in IPFS has increased during $2022$. Moreover, researchers report the \texttt{Agent Tesla} malware, which using phishing techniques, leads to an IPFS public gateway, disguising the download of malicious content. To better frame our research into the storage of the IPFS ecosystem, we also researched the file side. Our research is twofold, in the first case, we eavesdropped on the files requested by IPFS users, while in the second, more actively, we searched for files we randomly downloaded from well-known torrent sites.
\subsection{Bitswap Eavesdropping}
According to the operating rules of IPFS, when a user searches for a file, a one-hop inquiry is first performed through Bitswarm, requesting it from nodes with an active connection to the initiator. If none of them responds, the query is then served by the DHT. To collect data, we tweaked our node so that it maintains active connections with around $4000$ nodes; that is, according to our measurements, approximately $20$\% of the network's active nodes at that time. So when one of those nodes was looking for a file, thanks to Bitswap's functionality, that information would also go through us. This way, we could eavesdrop on about $20$\% of the network's requests and, in turn, request back to retrieve them. In total, we monitored the requests for $24$ hours while we set each request to last no more than $15$ seconds. This way, we avoided downloading very large files while, on the other hand, we cancelled the search in case it was routed through the DHT. In total, we collected $49155$ files with a size of about $13.7$ GB. To have a more complete picture of the type of files requested, we used the Python \texttt{mimetypes} module\footnote{\url{https://docs.python.org/3/library/mimetypes.html}} to find the MIME type of each file. We shall mention that it managed to classify $13691$ of the files. The latter can be attributed to Bitswap's design. When a user requests a file from Bitswap, the search is performed by the root CID of the file. The aforementioned file contains links to the chunks of which it is composed. Thus, when the requester receives the root CID and learns the CIDs of the chunks that make up the file, it requests through Bitswap consecutively all the chunks, which are essentially blocks of data. The file results illustrate that $3716$ are image files with MIME types ``image/png'', ``image/gif'', ``image/jpeg'', and 9148 are JSON files, which is the most common format for NFT metadata. The latter clearly demonstrates and confirms our initial statement that IPFS is a cornerstone of NFT data storage and Web3 in general. Among others, we fetched $177$ Javascript files and $27$ videos of type ``video/mp4''. We then fed the image files to the Python \texttt{Not Suitable For Work~(NSFW) Detector} module to determine whether IPFS is being used for inappropriate content. From the $1636$ image files it examined successfully, it found $33$ unsuitable \footnote{\url{https://pypi.org/project/nsfw-detector/}}. The above indicates that some users leverage IPFS' anonymity to host inappropriate content that is difficult for LEAs to track and take down.
\subsection{Torrent Files}
Very often, inappropriate files are found in the form of torrent files disseminated through torrent search engines. We downloaded a sample from various widespread torrent sites, ten popular torrents in total. We computed their CIDs locally to determine whether they are shared on the IPFS. This way, not only did we not add any illegal files to the IPFS network, but we also limited the possibility of tampering with the results of our upcoming searches. The ten different torrent files yielded $72$ different root CIDs. Each torrent file can contain a video file, a cover image for the video file, a text file with information about the file, etc. In turn, we made $72$ requests to the DHT for providers of these CIDs. We found providers for seven of them, and in fact, for most of them, more than one. The latter implies that IPFS users may also share the same content in torrents and that intellectual  infringement content is also distributed through IPFS.
\section{Countermeasures}
The amount of malicious nodes connected to IPFS is alarmingly high. Given the P2P nature of IPFS and its continuous exploitation, we believe that pruning nodes from the network might provide an initial measure of sanitising the network; otherwise, the benign peers facilitate the malicious ones. To this end, we opt for a periodical blacklist approach that is resolved through InterPlanetary Name System (IPNS). In essence, we propose the use of the proposed data crawling methodology to monitor the nodes on daily basis, the IPs are collected and using intelligence services, we determine whether the IP should be blocked or not. Each IP is four bytes long, so the expected size is rather small and easy to manage. For instance, using our experiments as a baseline, using the worst estimate of 32000 malicious nodes, the blocklist would be around 8KB if the IPs are directly stored. Given its size and possible optimisations (e.g. use binary search over the sorted list), searching whether the connected peers are malicious can be very efficient. Moreover, since the amount of nodes is tolerable, the collection of data from intelligence services can be rather fast. Of course, one could mask the IPs with the use of Bloom filter-based approaches  \cite{bloom1970space}. However,  since the IPs are very small in size, the corresponding filters would for sure be significantly bigger. Moreover, due to the probabilistic nature of bloom filters, there is always a margin of considering a benign node as malicious. That said, for our experiments, with an error probability of only 0.001\%, 93.6KB would be required, which is one scale of magnitude more.  

IPFS is becoming institutional, after all, there are many organisations participating in it and supporting it. Recent research efforts indicate that it could frame the existing banking system~\cite{kyc}, while at the same time it constitutes a cornerstone of Decentralised Finance~(DeFi). Our research does not intend to act as a brake on its use, on the contrary, it intends to inform, alert and promote its secure use. For instance, the network administrator of an organisation participating in the IPFS network can block the traffic towards and from a suspicious IP address by adding a rule to the firewall. Note that it can also remove alert fatigue from SOCs who might observe malicious IPs connected to the monitored infrastructure due to IPFS traffic. Finally, while IPFS provides the ability to disconnect from a node, it does not provide natively the option for the user to maintain a blacklist.

\section{Conclusions}
Open and decentralised systems are, by their very nature, prone to several attacks. However, given the crucial role of IPFS for Web3, it is essential to protect the ecosystem. Our measurements indicate that an alarming number of IPs reported as malicious through intelligence services are using IPFS. Rather than making it centralised, we opt for soft measures that allow nodes to isolate malicious ones selectively. We argue that this isolation can significantly benefit the network as the content of most of these nodes may be malicious, leading legitimate ones to facilitate nefarious acts and malicious campaigns. Therefore, their isolation, in the long run, may increase the robustness of the network and trust in it.

IPFS seems to have sacrificed part of the privacy to succeed in terms of performance, speed, and robustness~\cite{balduf2022monitoring}. This shortcoming can be exploited for malicious purposes, but it can also be leveraged by security analysts to monitor malicious nodes. Thus, apart from the fact that we can obtain critical information regarding a malicious node, such as its IP address, we can also monitor it from a content point of view, i.e., its requests as well as what it provides. Therefore, a future direction of this work is an extension of the implementation of the proposed filter so that it associates malicious nodes with the corresponding content.

\section*{Acknowledgements}
The authors would like to thank Dennis Trautwein for his insightful comments. This work was supported in part by the European Commission under the Horizon Europe Programme, as part of the project LAZARUS (https://lazarus-he.eu/, Grant Agreement no. 101070303) and the Horizon 2020 Programme, as part of the project HEROES (https://heroes-fct.eu/, Grant Agreement no. 101021801) and was also supported in part by the Research Center of the Athens University of Economics and Business.

The content of this article does not reflect the official opinion of the European Union. Responsibility for the information and views expressed therein lies entirely with the authors.

\end{document}